# First principles study of electronic and structural properties of single walled zigzag boron nitride nanotubes doped with the elements of group IV

Ali Bahari[a], Amir Jalalinejad[b], Mosahhar Bagheri[a] and Masoud Amiri[c*]

[a] Department of Solid State physics, University of Mazandaran, Babolsar, Iran
[b] Department of Molecular Sciences and Nanosystems, Scientific Campus, Via Torino 155, 30170 Venice Mestre, Italy
[c] Department of Physics, Shahid Beheshti University, G.C. Evin, Tehran 1983963113, Iran
[*] Corresponding author. E-mail address: ma_amiri@sbu.ac.ir

**Abstract**

In this paper, structural and electronic properties and stability of (10, 0) born nitride nanotube (BNNT) are considered within density functional theory by doping group IV elements of the periodic table. The HOMO-LUMO gap has been strongly modified and treated a dual manner by choosing B or N sites for dopant atoms. Formation energy calculation shows that B site doping is more stable than N site doping. Results also show that all dopants turn the pristine BNNT into a p-type semiconductor except for carbon-doped BNNT at B site.

**Keywords:** DFT; BNNT; electronic properties; stability

## 1. Introduction

BNNTs are considered by many scientists, because of their large band gap, excellent mechanical properties, strong piezoelectric features, chemical stability, great resistance to oxidation at high temperatures, independent band gap to geometry and hydrogen storage capability (in competition with CNTs, BNNTs would appear more successful in hydrogen storage) [1-10]. Rubio et. al. have predicted the existence of BNNTs for the first time [11]. Afterward, Chopra et al. have synthesized the multi-walled BNNTs by electric arc discharge [12]. Since then, more efficient methods have been developed to produce these nanotubes[13, 14]. The Band gap of BNNTs was measured theoretically and experimentally. Theoretical studies estimated its value to be around 4-6 eV [2, 4, 9, 12, 13, 15-18]. Experimental measurements based on Cathodoluminescence method have estimated values from 5.3 to 5.4 eV [19, 20]. Studying the effects of group IV elements on the electronic properties of BNNT's is an attractive subject that many workers have reported many of the structural and electronic properties for C, Si and Ge- doped BNNT in recent years [21-32]. However there are not any researches considering the effects of Sn and Pb. We applied an accurate method relative to recent researches, and our results are consistent with experimental data.

## 2. Computational details:

In this work, all restricted open shell electronic structure calculations are performed using GAMESS (US) package [33]. Macmolplot [34] and Avogadro packages[35] are used to prepare and visualize the data. Effective Core Potentials (ECP) as proposed by Stevence et al. [36, 37] along with d polarization functions are used as basis sets and B3LYP is considered as exchange-correlation functional [38, 39]. All structures are relaxed until maximum gradient reached less than 0.0001 Hartree/Bohr. An 11.6 angstrom length (10, 0) cluster BNNT including 60 boron and 60 nitrogen atoms saturated by 20 hydrogen atoms is considered as a pristine model (Fig. 1) and the radius of this model BNNT is 8 Å. To investigate the structural and electronic properties of X-doped (X=C, Si, Ge, Sn and Pb) BNNTs, each of these elements is substituted by a boron and nitrogen sites separately.

The formation energy for X doped is calculated using total energy calculations as the following equation[15, 40].

$$E_{Formation}^{B(N)}[X] = E_T^{B(N)}[NT+X] - E_T[NT] - \mu_X + \mu_{B(N)} \qquad (1)$$

Where $E_T^{B(N)}[NT+X]$ is the total energy of NT with one X atom. The upper indices refer to the site which X atom sits. $E_T[NT]$ is the total energy of the pristine system. $\mu$ is the atomic chemical potential of atoms (X or B or N) which is calculated as the total energy per atom of the $X_{10}$, $B_4$ and $N_2$ stable configurations.

3. **Results and Discussion**

Fig. 2 shows the resulting optimized structures of the (10,0) boron nitride nanotubes with C, Si, Ge, Sn and Pb at B and N sites. For all cases, local symmetry of the tube breaks up after relaxation and all dopants undergo a radial displacement outward the tube and form a tetrahedral structure with the three nearest neighboring atoms. Structural parameters are shown in Fig. 3 and are listed in table 1. As can be seen, bond lengths become longer as the atomic radius of impurities increase due to increment of atomic radius of impurities. However, bond lengths pertaining to impurity substitution at B site are shorter than impurities that are substituted at N site due to higher electronegativity of nitrogen than boron. The only exception is carbon: calculated bond lengths show that C-N bond length is shorter than that of pristine BNNT which could be understood by the fact that atomic radius of C (0.76 Å) is closer to N (0.72 Å) than B(0.84 Å) [40]. These results are consistent with bond angles listed in table 1. Bond length values for pristine, C-doped, Si-doped and Ge-doped BNNTs are in good agreement with previous studies [15, 25, 26, 28, 29, 31, 41], however, to our knowledge there are no theoretical and experimental results for Sn and Pb. In summary, table 1 shows that the deformation is smallest for carbon and is largest for Pb.

An important question to be addressed is the stability of the doped BNNT systems. The important quantity to compare stability of the above relaxed structures is their formation energies. Formation energies pertaining to the relaxed structures are listed in table 2. As can be seen, for carbon, substitution of nitrogen by carbon has lower formation energy with respect to carbon substitution at B site (with an energy difference by 2.78 eV). In other words, C impurity substitution at N site has higher stability. This result is in agreement with previous studies[41]. For Si-doped system, one can see that Si substitution at B site has lower formation energy than Si substitution at N site (by 1.66 eV) and hence Si substitution at B site has higher stability. This result is also in agreement with previous studies[25]. Similar trend can be seen for Ge, Sn and Pb. These results are also in agreement with structural parameters listed in table 1.

Studying the effects of impurities on electronic structure of doped BNNTs is another important issue. In table 3, HOMO-LUMO gaps of group IV-doped BNNTs are listed. As can be seen, the pristine (10,0) BNNT has a wide HOMO-LUMO gap of 5.94 eV (fig. 4 (a)). This value for HOMO-LUMO gap is in excellent agreement with experimental results [15, 42, 43]. A point to be mentioned here is that most DFT studies are based on Local Density Approximation (LDA) and Generalized Gradient Approximations (GGA) functionals that are well-known to underestimate HOMO-LUMO gaps, in particular for semiconductors. However, some studies have shown that hybrid functionals such as B3LYP and B3PW91 would lead to reasonable and even accurate results for band gaps [44, 45].

For carbon substitution at B site ($C_B$), (Fig. 4(b)), in-gap localized level is near the conduction band, this level is a donor level and the electron can be easily excited from this in-gap level to the conduction (empty) band. Carbon transfers a charge of 0.21e to the tube and hence the $C_B$-doped BNNT is an n-type semiconductor. For carbon substitution at N site ($C_N$), (Fig. 4(g)), in-gap localized level is an acceptor level and a charge of 0.9e is transferred from tube to the carbon impurity. These results are in agreement with previous studies[31].

For Si-doped BNNT system, substitution of Si at B site ($Si_B$), (Fig. 4 (c)), electrons from valence band can easily be excited to this acceptor level, indicating that the $Si_B$-doped BNNT system is a p-type semiconductor and a charge of 0.48e is transferred from the tube to the silicon impurity at B site. As for Si substitution at N site ($Si_N$)), (Fig. 4 (h)), Silicon acts as acceptor impurity and the resulting $Si_N$-doped BNNT is a p-type semiconductor. The charge transferred from tube to the silicon impurity is 0.049e. This small amount of transferred indicates that Si impurity doped at B site acts as a deep impurity center.

Ge and Sn exhibit similar behaviors to silicon, respectively, when they are substituted at B site (Fig. 4 (d) and Fig. 4 (e)), the induced level shifts to the valence band, meaning that $Ge_B$-doped and $Sn_B$-doped BNNT systems exhibit p-type semiconductor characteristics. As for $Ge_N$- and $Sn_N$-doped BNNTs, the induced level exhibits similar trends to $Ge_B$-doped and $Sn_B$-doped BNNTs, $Ge_N$- and $Sn_N$-doped BNNTs also exhibit p-type semiconductor characteristics.

However, the situation is quite different for Pb: when Pb is substituted at B site ($Pb_B$), no in-gap state is induced within the HOMO-LUMO gap (Fig. 4 (f)) which can be understood as follows: when Pb is substituted at B site, the induced energy level, with an energy level of -6.31 eV, lies on top of the valence band, causing the top of valance band to be partially filled, and therefore changing the insulator behavior of pristine BNNT to p-type semiconducting behavior with a reduced band gap of 5.33 eV and results in a charge transfer of 0.67e from tube to the Pb impurity. Similar theoretical study on the effects of gold doping have reported similar behavior to $Pb_B$ case [15]. For Pb substitution at N site ($Pb_N$), an in-gap energy level appears in the pristine gap, with energy of -3.94 eV (fig. 4 (k)), about 1.15 eV above the valence band and hence exhibiting a p-type semiconductor characteristics and a charge of 0.26e has been transformed from tube to the Pb impurity.

4. Conclusion

In summary, based on DFT calculations, we have investigated the structural and electronic properties of group IV-doped BNNTs. All resulting doped BNNT (except for $C_B$) systems exhibit p-type semiconductor characteristics. However, $C_B$ shows n-type semiconductor behavior. The $X_B$ (X=C, Si, Ge, Sn and Pb)-doped BNNTs (except for carbon) have lower formation energies than $X_N$-doped BNNTs meaning that $X_B$-doped BNNTs have higher stability. According to our calculations, the C doping has a good stability and the formation energy of p type is so far from to n type, therefore the production mechanism can be easily achieved. One could control the temperature or the energy of carbon source in a suit low range to produce p type semiconductor. On the other hand by increasing the energy of carbon energy in a suit order we could obtain n-type semiconductor.

Table 1 Structural parameters of BNNT. The numbers identifying the bond lengths and angles refer to Fig. 2.

|  | B Site | | N Site | |
| --- | --- | --- | --- | --- |
|  | Bond length(Å) | Angle(deg) | Bond length (Å) | Angle(deg) |
| C | 1-2=1.44<br>1-3=1.45<br>1-4=1.45 | 2-1-4=117.0<br>2-1-3=117.0<br>3-1-4=113.5 | 1-2=1.54<br>1-3=1.55<br>1-4=1.55 | 2-1-4=118.2<br>2-1-3=118.2<br>3-1-4=114.1 |
| Si | 1-2=1.73<br>1-3=1.75<br>1-4=1.75 | 2-1-4=106.0<br>2-1-3=106.0<br>3-1-4=105.8 | 1-2=1.96<br>1-3=2.1<br>1-4=2.01 | 2-1-4=97.2<br>2-1-3=97.2<br>3-1-4=91.4 |
| Ge | 1-2=1.81<br>1-3=1.85<br>1-4=1.85 | 2-1-4=103.2<br>2-1-3=103.2<br>3-1-4=102.9 | 1-2=2.00<br>1-3=2.06<br>1-4=2.06 | 2-1-4=95.3<br>2-1-3=95.3<br>3-1-4=89.5 |
| Sn | 1-2=1.98<br>1-3=2.03<br>1-4=2.03 | 2-1-4=96.2<br>2-1-3=96.3<br>3-1-4=97.1 | 1-2=2.12<br>1-3=2.22<br>1-4=2.24 | 2-1-4=86.9<br>2-1-3=87.1<br>3-1-4=80.5 |
| Pb | 1-2=2.33<br>1-3=2.20<br>1-4=2.20 | 2-1-4=78.7<br>2-1-3=78.8<br>3-1-4=85.2 | 1-2=2.42<br>1-3=2.43<br>1-4=2.36 | 2-1-4=73.8<br>2-1-3=67.1<br>3-1-4=75.6 |
| Pristine | Distance(Å) | | Angle(deg) | |
|  | 1-2=1.46<br>1-3=1.46<br>1-4=1.46 | | 2-1-4=120.0<br>2-1-3=120.0<br>3-1-4=119.3 | |

Table 2  Formation energies of X-doped BNNT on B or N sites.

|  | $E_{Form}$(eV) B Site | $E_{Form}$(eV) N Site |
|---|---|---|
| C | 4.58 | 1.79 |
| Si | 4.06 | 5.73 |
| Ge | 5.71 | 6.32 |
| Sn | 6.79 | 7.30 |
| Pb | 6.99 | 7.34 |

Table 3 HOMO, LUMO and HOMO-LUMO gap of x-doped and pristine BNNT.

| | B Site | | | N Site | | |
|---|---|---|---|---|---|---|
| | HOMO (eV) | LUMO (eV) | HOMO-LUMO (eV) | HOMO (eV) | LUMO (eV) | HOMO-LUMO (eV) |
| C | -2.34 | -0.70 | 1.64 | -5.06 | -0.70 | 4.35 |
| Si | -3.80 | -0.68 | 3.12 | -4.73 | -1.14 | 3.59 |
| Ge | -4.62 | -0.68 | 3.94 | -4.73 | -1.11 | 3.61 |
| Sn | -5.38 | -0.65 | 4.73 | -4.84 | -1.25 | 3.59 |
| Pb | -6.31 | -0.97 | 5.33 | -3.94 | -1.38 | 2.55 |
| Pristine | HOMO (eV) | | LUMO (eV) | | HOMO-LUMO | |
| | -6.63 | | -0.69 | | 5.94 | |

Fig. 1 The relaxed pristine structure of BNNT all dangling bond was saturated by hydrogen.

Fig. 2 The relaxed doped structure of BNNT. (a) C, (b) Si, (c) Ge, (d) Sn and (e) Pb doped at B Site. (f) C, (g) Si, (h) Ge, (i) Sn and (j) Pb doped at N site.

Fig. 3 Structural parameters for (a) X doped at B Site, (b) X doped at N site Where X is C, Si, Ge, Sn and Pb.

Fig. 4 Density of states and energy levels for (a) pristine, (b) C, (c) Si, (d) Ge, (e) Sn and (f) Pb doped at B Site and (g) C, (h) Si, (i) Ge, (j) Sn and (k) Pb doped at N Site

Fig. 1

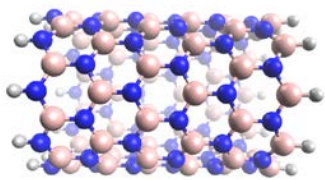

Fig. 2

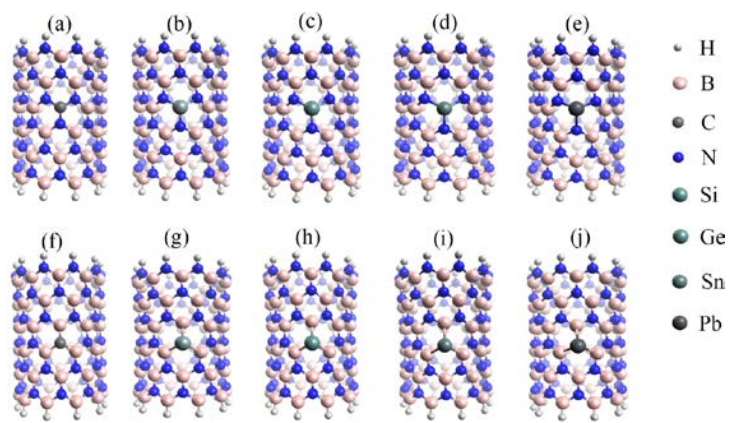

Fig. 3

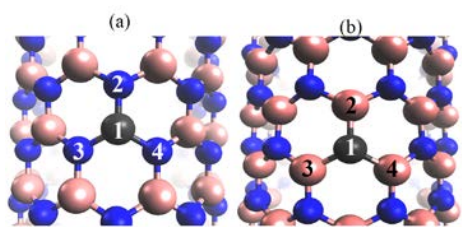

Fig. 4

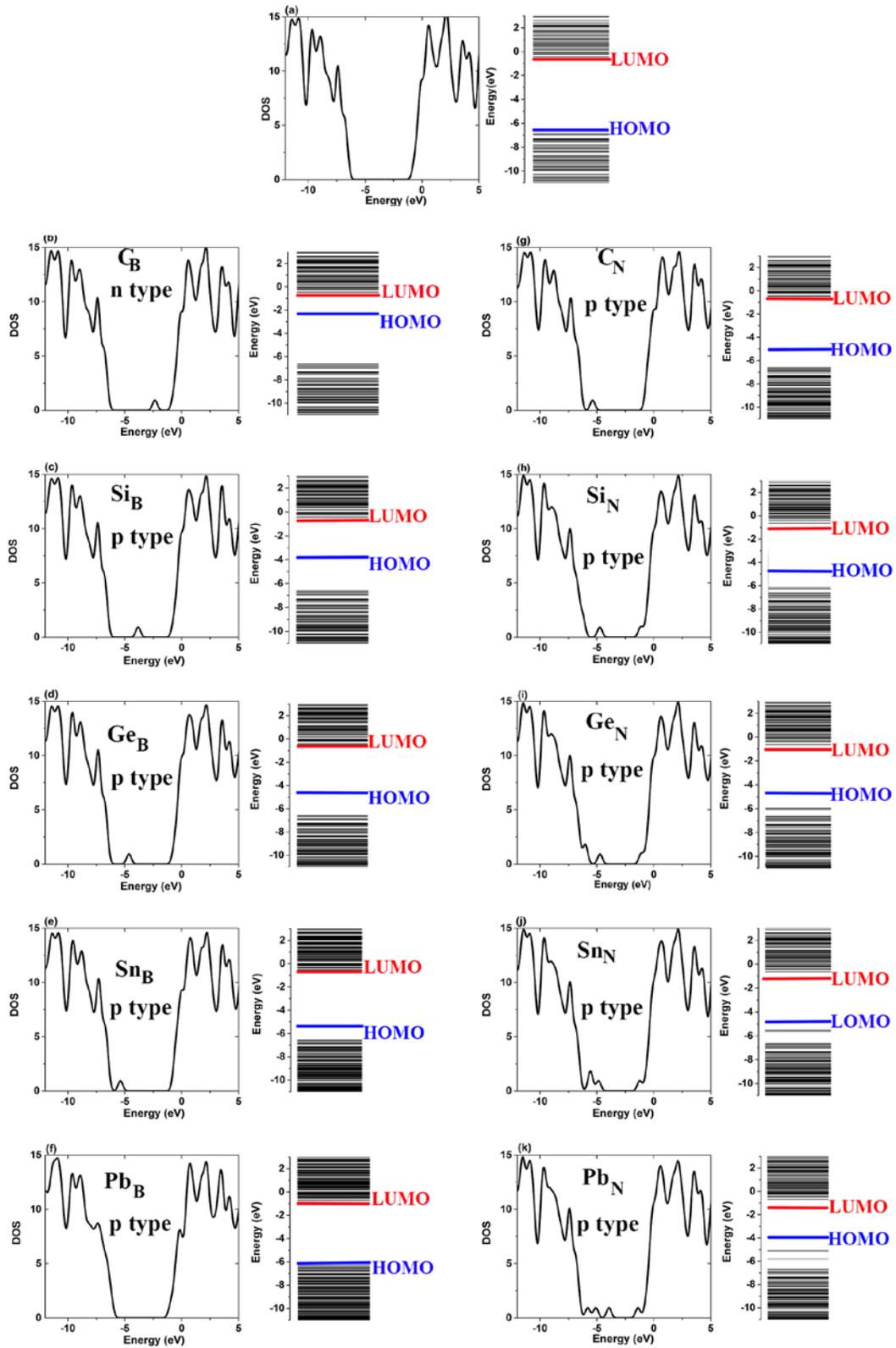